\documentclass[conference]{IEEEtran}
\IEEEoverridecommandlockouts
\usepackage{cite}
\usepackage{amsmath,amssymb,amsfonts}
\usepackage{algorithmic}
\usepackage{graphicx}
\usepackage{textcomp}
\usepackage{xcolor}
\def\BibTeX{{\rm B\kern-.05em{\sc i\kern-.025em b}\kern-.08em
    T\kern-.1667em\lower.7ex\hbox{E}\kern-.125emX}}
\begin{document}

\title{At the Edge of a Seamless Cloud Experience
\thanks{This work has been made with support in partnership with Proximus Luxembourg S.A.}
}

\author{\IEEEauthorblockN{Samuel Rac\IEEEauthorrefmark{1} and Mats Brorsson\IEEEauthorrefmark{2}}
\IEEEauthorblockA{\textit{Interdisciplinary Centre for Security, Reliability, and Trust} \\
\textit{University of Luxembourg}, Luxembourg \\
Email: \IEEEauthorrefmark{1}samuel.rac@uni.lu, \IEEEauthorrefmark{2} mats.brorsson@uni.lu}
}

\maketitle

\begin{abstract}
There is a growing need for low latency for many devices and users. The traditional cloud computing paradigm can not meet this requirement, legitimizing the need for a new paradigm. Edge computing proposes to move computing capacities to the edge of the network, closer to where data is produced and consumed. However, edge computing raises new challenges. At the edge, devices are more heterogeneous than in the data centre, where everything is optimized to achieve economies of scale. Edge devices can be mobile, like a car, which complicates architecture with dynamic topologies. IoT devices produce a considerable amount of data that can be processed at the Edge.

In this paper, we discuss the main challenges to be met in edge computing and solutions to achieve a seamless cloud experience. We propose to use technologies like containers and WebAssembly to manage applications' execution on heterogeneous devices. 

%The experience of managing applications at the Edge should be as easy as it is in the cloud.  

\end{abstract}

\begin{IEEEkeywords}
Cloud computing, Edge computing, Container orchestration, WebAssembly
\end{IEEEkeywords}

\section{Introduction}

Cloud computing is evolving and so are the requirements that modern applications put on the infrastructure and services provided. On the application side, we particularly see the need for Edge computing growing rapidly. The definition of the \emph{edge} varies in the literature. From a cloud provider perspective, the point-of-presence that a user connects to is the edge. In a cyber-physical system (CPS), the edge is likely to be the system to which IoT devices are connected. For the purpose of this paper, we define \textit{Edge computing} to be application processing that takes place outside a  data centre. Some people also call this \textit{Fog computing}~\cite{AllYouNeedToKnowCompleteSurvey}.

\begin{figure}
  \centering
    \includegraphics[width=\columnwidth]{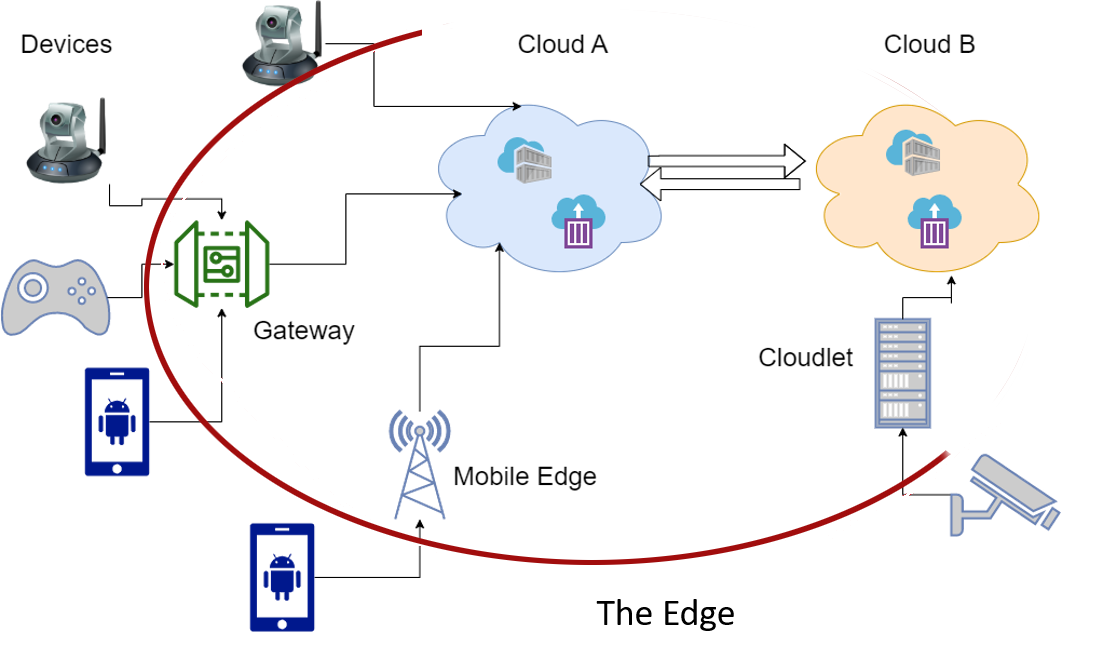}
  \caption{A hierarchical cloud infrastructure.}
  \label{hierarchical_cloud}
\end{figure}

An infrastructure able to support different kinds of edge applications (given the above definition of the Edge), might be quite complex. IoT, and other, devices might be connected to the data centres directly or through gateways acting as aggregators of device data. Some connections are directly through the internet, others through other means like 5G/LTE or edge radio protocols. In any and all of these locations, devices, gateways, telecommunication base station and cloud nodes, there might be compute and storage capabilities, and we then have an infrastructure from device to network edge, to cloud edge and to central cloud data centre with many steps in between, each possibly having compute capabilities. See Figure~\ref{hierarchical_cloud}, which illustrates one possibility for this infrastructure architecture. 

Irrespective of what applications are run on such an infrastructure, the current software deployment model is very fragmented and highly manual. Assuming the situation in Figure~\ref{hierarchical_cloud}, software may be deployed at several levels:

\begin{itemize}
\item	At a data centre in the cloud (where there is a choice of multiple
        points-of-presence).
\item	At a data centre of another cloud provider.
\item	At a gateway node.
\item	At a network edge node.
\item	At an edge device.
\end{itemize}

The state-of-the-art method to deploy an application consisting of multiple components, a distributed system, is to use a container orchestration method such as Kubernetes. This allows for a descriptive model of specifying the desired state of the application and Kubernetes will work at making sure that the current state matches the desired state even during disturbances such as node failure and changes of requirements such as sudden surge in traffic.

This model works well for an application in one, or spanning a few, data centres in one cloud. However, if you want to deploy an application spanning multiple or all levels of a hierarchical cloud according to Figure~\ref{hierarchical_cloud}, all levels outside the traditional cloud are mostly managed manually without any adaptation on the changing needs of an application. 

Furthermore, while data centre nodes tend to be quite homogeneous, gateways and edge devices come in many shapes and forms and may use different processor architectures (currently mainly x86 or ARM). Currently, you need to explicitly target the architecture of the node where the code should run. Finally, there is another level of heterogeneity: accelerators. In data centres there are GPU-accelerators, often used for training machine learning models. Increasingly we see other types of accelerators as well, such as FPGAs and Tensor Processor Units (TPUs). At the edge, you often see many kinds of accelerators for machine learning inference, graphics processing, etc. 

All-in-all, developing for this heterogeneous infrastructure landscape is difficult and may lead to cloud vendor lock-in and infrastructure over-provisioning as well as inferior performance and/or power consumption characteristics. We believe deploying applications to an Edge application should be as easy as deploying an application to services like Google GKE Autopilot~\cite{gke_autopilot}.

In this paper, we explore the state of the art in seamless cloud computing across such an infrastructure along three dimensions: i) architecture, ii) execution, and iii) orchestration. The \textit{Architecture} dimension is about technologies and concepts on how to connect components to form a seamless cloud experience. The  \textit{Execution} dimension is about various solutions to deal with heterogeneous platforms, and, finally, the \textit{Orchestration} dimension is about solutions to manage an application spreading across the infrastructure. 

\section{Architecture}

Finding an architecture for a seamless cloud computing experience is not an easy task. 
The challenge here is to identify the different layers and aggregate them to form a seamless cloud experience. We review here the main architectural principles that has been proposed to extend traditional cloud architectures to the edge and in between. 

\subsection{Topology}

%\subsubsection{Star topology}

\subsubsection{Cloudlet topology}

Cloudlet is an essential paradigm to achieve a seamless cloud. Cloudlet can be a server, a computer, or a little cluster located at the edge of the network, typically at one hop of the devices. With mobile devices and Cloud, they achieve a 3-tier continuum architecture \cite{AllYouNeedToKnowCompleteSurvey}. Cloudlets are designed to provide services to mobile devices \cite{EdgeSurvey}. The key idea of Cloudlet is to enable mobile devices offloading in their vicinity. On these devices, energy consumption is one of the major parameters for offloading. Cloudlets must be close to mobile devices to satisfy the  Quality of Service (QoS).
Cloudlets should also be able to work without the Internet. Working in a local area would also avoid congestion in the core network.
To have a seamless cloud experience, Cloudlet, traditional Cloud, and Mobile devices have to work in synergy. Offloading on Cloudlet or traditional Cloud should be transparent for the end-user. Choosing one or the other must consider the following parameters: latency, task size, energy, resource availability.

\subsubsection{Ad-hoc topology}

Ad-hoc edge computing aims to create a dynamic, distributed and decentralized infrastructure to take into account edge constraints \cite{JUANFERRER2021548}. Ad-hoc edge is a very resilient architecture that relies on a decentralized network, this should help avoid having a single point of failure. Ad-hoc network should have a lightweight implementation not to disturb the normal functioning of edge devices\cite{AdHocEdgeCloud}.

Mobility is a game-changer for the practices of resources management. When devices change their location, this causes problems: unreliable connection, resource availability and intermittent connection to the internet. Mobility affects the availability of edge nodes. 
Dynamicity present in resource availability, also named node churn,  reflects the behaviour of resources that appear and disappear from the system\cite{AdHocEdgeCloud},\cite{JUANFERRER2021548}. Also, the availability of devices can be determined by their energy management and battery discharge. 

\subsection{Networking}

Networking is a keystone in the construction of a seamless cloud. A reliable networking architecture is required for running distributed applications. The development of paradigms like micro-services architectures also shows dependences on the network.
In addition, networking devices can have an active role in edge computing. In-network computing is a paradigm where applications are distributed through the network infrastructure
\cite{COOKE2020395}. E. g., a switch or a router should become a small edge node and propose computing power to run small applications.

%\subsubsection{SDN}

Software-Defined Networking (SDN) is a technology that can provide seamless networking configuration. Configuring networking takes time, there are devices to configure manually (e. g., routers, firewalls). SDN permits to automate networking configuration and management. The idea of SDN is similar to Operating System, adding abstraction on top of hardware to simplify uses. SDN has a separation between the control plane and the user plane to increase network management flexibility.
Edge devices can be mobile, having an intermittent connection or limited bandwidth. Therefore, it is important to easily manage networking to be able to adapt to this dynamic networking topology.

SDN flexibility allows having a service-centric architecture~\cite{EdgeBenefitsFromSDN}. Networking can be dynamically adapted to the task's needs. The SDN controller has a high-level view of the network that helps deploy an application and its networking. If the application needs to scale up or down, networking will dynamically follow its needs.

%SDN flexibility is important to achieve mobility of the end-user devices. For example, if a mobile device goes from an access point to another, routing will be dynamically adapted. If this device has offloaded a task to Cloud or Cloudlet, the results will be delivered at its new position directly; this will avoid unnecessary traffic.

%\subsubsection{Networking virtualization}

A goal of edge computing is to provide low latency and location-aware services. The idea of a seamless cloud experience is to execute tasks at the edge of the network as if they were processed in the traditional Cloud, with a low latency benefit. When tasks or data goes to the Cloud through the edge node, routing should not affect performances.
SDN and Networking virtualization provides the flexibility needed for a seamless cloud experience.

Besides, edge computing will take advantage of the fifth generation of mobile networks (5G). It seems difficult to achieve mobility and use cases like V2X without using those networks. The 5G network will bring three major innovations: enhanced Mobile Broad Band (eMBB), Ultra-Reliable Low-Latency Communications (URLLC), and massive Machine-Type Communications (mMTC). 5G technologies rely in part on the virtualization of networking.

SDN will make networking easier to deploy for telco operators. There will be less need for specific and expensive hardware thanks to virtualization. Also, resources will be mutualized for better performance.

Network slicing is a 5G feature that proposes a new way of virtualization for network bandwidth usage. It allows a more dynamic and flexible infrastructure. Network slices can be design for a particular purpose, e. g. reserve resources to achieve ultra-low latency on one slice and manage many users on another. Slices can guarantee some levels of Quality of Service (QoS). SDN has a role in network slicing deployment.

% Those technologies are also important for being multi-vendor and multi-tenant. Seamless Cloud will involve various actors at many stages.  Telco operators also need to support roaming features.

%SDN, VNF, and MEC are inseparable and required to achieve the goal of ultra-low latency.

Virtualization is more and more important in networking. With Network Function Virtualisation (NFV), network function (NF) can be managed like a VM. According to \cite{KAZEMIFARD2021107809}, NF can be containerized, and VNF should become Containerized Network Functions (CNFs). Containerization allows even more flexibility and enables Kubernetes-based networking.

%Virtualization can even be used in the Radio Access Network (RAN), with virtualized RAN (vRAN). This is also based on splitting the control plane and the user plane to focus on services. Open RAN initiatives were created to avoid vendor lock-in and develop multi-tenancy cooperation. The O-RAN Alliance is one of the major contributors to Open RAN initiatives. Virtualization of RAN goes one step further with cloud RAN (cRAN). The idea is to distribute RAN functions on cloud infrastructure, using cloud-native tools and practices. cRAN brings cloud flexibility to the telco universe.

%Edge computing and Networking are progressing in synergy. Networking is essential to edge computing development, and Cloud paradigms are used to develop network performances and abilities.

\section{Execution}
The purpose of Edge computing is to transfer computation abilities at the edge of the network. Systems can be composed of many kinds of devices. It then becomes necessary to find a way to manage all of them.
Managing a cluster with a lot of similar nodes is a problem with a known solution. However, it becomes problematic when we need to manage heterogeneous nodes. In a heterogeneous cluster, there are multiple computing platforms; the challenge is to address all of them.

There are many ways to address most of those computing platforms. However, we will not discuss virtual-machine (VM). VM is a significant paradigm in cloud computing, but it often requires too much computation power for small devices usually found at the edge. Then, we focus on lighter virtualization technology like containers.

WebAssembly is also a technology to consider in an edge computing environment. It brings lightweight and fast execution, like containers, but in a more secure way by providing sandboxing and with architectural independence. Security is fundamental in a multi-tenant context where providers execute untrusted codes on their platforms.

\begin{figure}[b]
  \centering
    \includegraphics[width=\columnwidth]{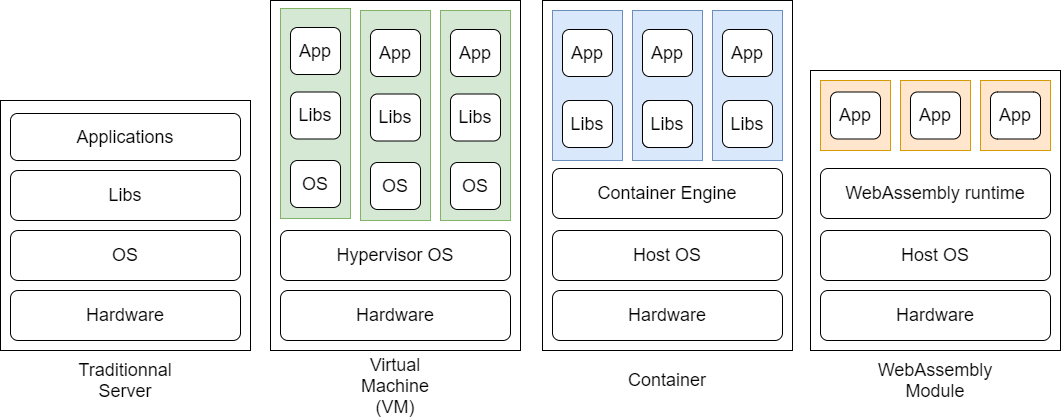}
  \caption{Comparison of architecture}
  \label{architectures}
\end{figure}

Figure \ref{architectures} compares WebAssembly’s architecture to other models found in Cloud Computing.

\subsection{Containers}
Containers is rapidly becoming the standard deployment technology in cloud computing. They allow better management of resources and make the development/deployment of applications easier. Also for Edge computing, container technology can be considered.

According to Hong and Varghese, \cite{ResourceManagement}, containers propose good performance for edge computing. However, they are not ready to deal with heterogeneity. They do not support specific hardware like graphic processing units (GPU), data processing units (DPU), tensor processing units (TPU), or field-programmable gate array (FPGA). Those hardware accelerators are essential to achieve ultra-low latency on specific applications at the edge.
However, building container for multiple targets requires additional steps. Containers strongly depend on their host; e.g., a Linux image cannot be directly executed on a Windows host.

Kata containers \cite{KataContainersforEnablingMEC} are really lightweight VM that can be managed like a container (e.g., in Kubernetes). Using them allows having workload isolation and better security. Combining VM and containers could be a solution to manage hardware accelerators and security.

\subsubsection{OCI} OCI\footnote{https://opencontainers.org/} propose standards for container formats and runtimes. It allows simpler management of container images (stored in registries). Docker containers, Kata containers, and even WebAssembly modules are compatible with this structure. In order to manage heterogeneous nodes at the edge, following OCI standards will help to deploy applications.

\subsubsection{Live migration} Live migration between edge devices is an important mechanism. Some devices at the edge can be mobile, and their applications have to follow them to always be as close as possible. Xu et al~\cite{sledgeLiveMigration} present a tool for Docker container live migration. They explain that container migration is more complicated than VM migration. Docker container migration needs to migrate image, runtime, and context.

\subsection{WebAssembly}
WebAssembly (also called WASM) is a new technology design for web applications that now rapidly gains interes outside the web. The idea behind WASM is to create a new assembly standard that proposes a unique binary that can run on every processor with good performance. WASM also comes with a memory-safe structure, lightweight sandboxed execution that allows a high-security level and an efficient execution. It is compatible with most high-level programming languages (e. g. Rust, C/C++, ...).

WebAssembly should have a significant contribution to edge computing. This improvement should allow developers to maintain a single code-base and address most devices with a single binary. Considering that edge devices will have different architecture, WebAssembly should become a game-changer. WebAssembly applications should be deployed on every edge device seamlessly without worrying about the platform or architecture.

The major web browsers (Chrome, Edge, Firefox, and Safari) now support WebAssembly. Apart from browsers, WASM can run with many different runtimes.
 
WebAssembly Standard Interface (WASI)\footnote{https://wasi.dev/} aims to propose standards to facilitate communication between WebAssembly conceptual machine and its host system. WASI has a goal quite similar to that of POSIX.

Wasmtime is a runtime that implements the WASI standards. It is still under development as WASI is not yet completed. Wasmtime is, therefore, still limited in its uses because of missing networking functionalities. However, implementations are already tested. It is only a matter of time before the standardization is complete.

The flexibility and performances of WebAssembly are strongly linked to its runtime. WebAssembly can be interpreted; it supports more platforms but has lesser performance. Some runtimes support Just In Time Compilation (JIT), performances are better for heavier tasks, but binaries are larger. Then, Ahead-Of-Time (AOT) compilation provides the best performances. Execution performance really depends on the runtime because they do not have the same level of optimization. Runtimes evolve quickly and often, therefore, it is not so easy to evaluate WASM performance.

\subsubsection{Running WebAssembly everywhere}
WASM binary can be executed on almost every architecture (x86, x86\_64, ARM, RISC-V, PowerPC, MIPS, Xtensa, ARC32, ...) and platform (laptop, smartphones, SBC like Raspberry Pi, MCUs like Arduino, ...) thanks to runtimes like wasm3\footnote{https://github.com/wasm3/wasm3}. This wide support means that the same software can run on almost every device at the edge, even with limited resources. The task of managing resources becomes easier with WASM. In a world where most data will be collected and processed at the edge, managing a heterogeneous swarm of devices will be simpler when executing the same binary.

Most high-level programmation languages can target WebAssembly. However, it is still challenging to uses some.  It is tricky to implement Garbage collectors when the available memory is set before the execution. Languages like Java cannot be used with WebAssembly yet. However, those languages with a high-level virtual machine do not seem to be good candidates for edge computing. They are not optimized enough for edge device constraints like memory or CPU.

\begin{figure}[b]
  \centering
    \includegraphics[width=\columnwidth]{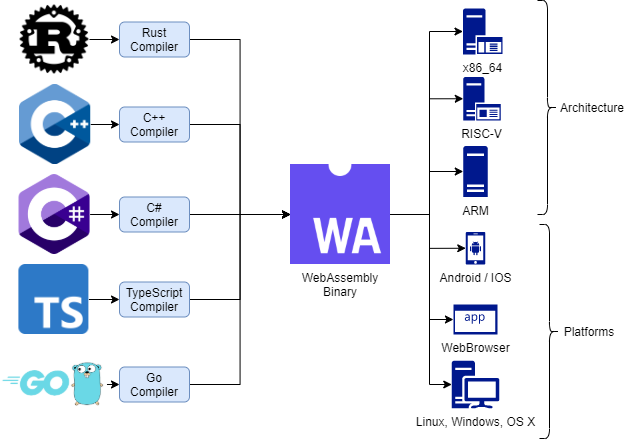}
  \caption{WASM}
  \label{Wasm}
\end{figure}

Figure \ref{Wasm} shows the different platforms and architectures that can be targeted.

\subsubsection{Krustlet}
Krustlet is a tool that allows WebAssembly modules to run on Kubernetes nodes. The edge environment is fundamentally heterogeneous; it is essential to have this in mind to deploy a cluster at the Edge. Krustlet permits using the state of the art of orchestrator at the Edge. WebAssembly modules can be run on every node, which enables to have a cluster with heterogeneous nodes.
Krustlet implements the Kubelet API, the pod manager present on every node in a Kubernetes cluster (the pod is the atomic unit manageable on Kubernetes). WebAssembly modules can be pushed in a registry following the OCI standards. Then WASM modules can be scheduled on krustlet nodes. Nodes that run Krustlet can execute natively WebAssembly modules, using WASI runtime (wasmtime) by default. 
Currently, this project is still limited by the progression of WASI, especially for networking missing features.

\subsubsection{Wasmcloud}
Wasmcloud\footnote{https://wasmcloud.dev/} is an open-source platform for developing applications in WebAssembly and deploy them to the Cloud. The ability to deploy a WebAssembly module on a cluster is important for edge computing, where nodes are heterogeneous. The project wants to improve application development in a faster and safer way with less boilerplate code. 
Wasmcloud wants to facilitate micro-service deployment with the actor model. In wasmcloud, an actor is the smallest deployable unit. Every actor is a small WebAssembly module that can handle messages from a capability provider. Capability providers are dynamic libraries that provide interfaces like an HTTP server, access to databases, or storage to an actor. Actors are reactive, and they react to messages they receive. They are secure because of privileged-based security and the inherent advantages of WebAssembly.
With the help of Krustlet, Wasmcloud actors’ can be deployed into a cluster at the edge.

\subsection{Accelerators}

Edge computing has to manage a growing demand in data processing while reducing latency. CPU alone can not meet this increasing demand. Hardware accelerators respond to this challenge, they are circuits designed for a specific use case. E.g., accelerators can be designed to handle data for the V2X use case in real-time. In their field of applications, hardware accelerators are really better than CPU and can consume less energy. It is also possible to virtualize accelerators to make the most of their capabilities \cite{COOKE2020395}, virtualization brings flexibility in utilization.
GPU, FPGA, and ASIC are the main accelerators for edge computing. They can be used together with CPU to form heterogeneous platforms \cite{CPUFPGAHeterogeneousArchitecture}.

\subsubsection{GPU} 
Graphics processing unit (GPU) are circuits design to execute parallel applications faster. They were initially designed to process images and are used in a multitude of applications like AI. They can be found everywhere: in data centers, personal computers, or embedded devices. GPUs are the more common accelerators, they also are the more energy-consumer. Frameworks like OpenCL or CUDA can be used to program GPU.

\subsubsection{ASIC}  
Application-specific integrated circuit (ASIC) are circuits designed for a specific purpose. They propose high performances and security for small size and lower power consumption. \cite{ASurveyonEdgeComputingSystemsandTools}. ASICs are the best alternative for a very specific task, but the flip side of the coin is their lack of flexibility. They are totally incapable of doing anything different than what they were designed to do.

\subsubsection{FPGA} 
Field-programmable gate array (FPGA) are circuits designed to be reconfigured after manufacturing. Reconfiguration can be static, i.e.,  before program execution, or dynamic, i.e., during runtime. Reconfiguration can be partial or global depending on the circuit \cite{CPUFPGAHeterogeneousArchitecture}. They are designed to achieve specific tasks like ASICs, but they are more versatile and can be reconfigured for many other usages. FPGAs provide low latency and low power consumption that make them suitable for edge or embedded devices. Public cloud providers propose access to FPGA, analogous access could be available for edge computing. FPGA are secure devices from an external perspective, but giving access to reconfigurable hardware in a multi-tenant environment could raise security issues. FPGAs’ reconfigurability can be very useful for a scheduler at the edge. If a computing unit is missing or not available, an FPGA can be dynamically reconfigured to execute a new task. Hardware description language (HDL)  like VHDL can be used to configure FPGAs.

\section{Orchestration and management}

Edge computing will extend the cloud with many different devices. The challenge will be to make every device work together. For this, orchestration methods from cloud computing need to be extended for edge computing. Edge computing orchestration needs to take into consideration the volatility and the heterogeneity of resources. Tools like Kubernetes should be improved to support the edge computing challenges. In order to manage edge complexity, automation methods should be used.

\subsection{Scheduling}

Scheduling at the edge is a fundamental challenge. There is a huge number of parameters to take into account. 
In addition, this heterogeneous distributed architecture will be multi-tenant.

Kubernetes has become a standard in scheduling. However, some requirements are needed in a heterogeneous cluster. Kubernetes has to manage the different node’s architecture. In addition, hardware accelerators have to be taken into account. In order to have a seamless cloud, a heterogenous cluster has to identify its different capabilities, and developers have to add constraints to their applications. A good scheduler has to ensure that application constraints meet the cluster and node capabilities. Other parameters like energy consumption should also be taken into account by the scheduler. Auto-scaling and auto-discovery mechanisms should also be addressed.

There is, therefore, a need for standardization to achieve scheduling at the edge. The container standard defined by the OCI could be a good candidate. As explained in the previous section, most of the containers support it, like Docker or WebAssembly. Having a standardized container will allow having an atomic unit to manageable by the scheduler. It is important to provide resource abstraction (i.e., CPU architecture, hardware accelerator) to permit more efficient scheduling by removing a significant constraint. It is necessary insofar as devices are more and more heterogeneous.

This atomic unit should include information to consider the following parameters: available resources, need of bandwidth, low latency requirement, real-time execution, the energy available, device location or context, hardware accelerator. 

Another issue of such a scheduler is being integrated into a context of cooperation between edge and Cloud. A distributed application may need to execute some tasks at the edge, while others may need the powerful computing capabilities of the Cloud.

\subsection{Multi-access Edge Computing}

Multi-access Edge Computing (MEC) is a new paradigm that aims to provide cloud abilities at the network's edge.
MEC also permits application offloading and data aggregation~\cite{giust2017multi}.
ETSI\footnote{https://www.etsi.org/technologies/multi-access-edge-computing} defines many aspects of MEC and presents some use cases. MEC can be used to achieve ultra-low latency, high bandwidth, and Real-Time performance at the edge. These technologies enable many applications like gaming, Virtual Reality (VR), Augmented Reality (AR), Internet of Things (IoT), or Vehicle-to-everything (V2X). With MEC, tasks can be offloaded from small devices to more powerful ones. Energy is another essential criterion for offloading. A low battery device can offload a task to save battery.

MEC is a key enabler for 5G technologies because of its position at the edge of a network. MEC can be deployed in a Radio access network (RAN) and provides computation and storage capabilities. Data generated by edge devices can be collected and processed directly in MEC. This can provide better privacy and security of personal data. Data are not systematically sent to a central authority but processed locally. Processing data at the edge will help to unclog the network and data centers. RAN and MEC can enable context-aware services. 
Taking into account devices and users' mobility is essential, e. g. a connected smart car needs to process its data locally to have ultra-low latency. While a vehicle is a mobile device, data processing tasks should migrate from one node to the devices' closest. MEC has to be able to realize the live migration of tasks to go with devices. 

Larger computing abilities at the edge can be used to train Artificial Intelligence (AI) and Deep Learning (DL).
All data required for this task can be found directly at the source. Those data should help to have better models while saving bandwidth to the data centers. Using the models during the inference phase should also be done at the edge.
Managing and scheduling tasks while considering all of those criteria in this heterogeneous environment is a significant research field. Scheduling algorithms should find the better node at the better location to improve MEC performance.

MEC has a key role in edge computing. It is a link between Cloud and telco.

Kubernetes should be extended to be usable at the edge. In this case, MEC would become an interesting use case. 
However, it is necessary to package the application in a container-friendly format in order to manage it.
Projects like KubeEdge\footnote{https://kubeedge.io/} proses this extension. KubeEdge has a control plane on the
cloud side and worker nodes on the edge side. This tool also considers network constraints like unreliable connections to allow edge nodes to continue to work even in the case of networking failure.

\subsection{Serverless}

Serverless is a paradigm where a cloud provider offers an abstraction of the resources allocated to a client. 
According to \cite{letaureau}, serverless is a technology that is increasingly being used. Developers only have to pay attention to the code they wrote; resources and infrastructure will be dynamically provided according to the need. It is easy to use with the support of most high-level programming languages. In contrast to a dedicated server, 
serverless is cost-efficient; only used resources are billed. However, serverless is less suitable for long-running tasks. When execution times become too long, serverless become more expensive. Dedicated servers 
are better candidates for long-run resource-intensive execution.
Serverless is an event-driven paradigm; an external event triggers the application execution. Apart from this case, no resources are consumed.

Serverless can be seen as the addition of Function as a Service (FaaS) and Backend as a Service (BaaS).
FaaS is a service that permits to trigger code execution by events. Execution is stateless and needs external storage to save data.
BaaS are services needed for FaaS, e.g., external storage, database, networking functions...

A common practice in serverless is Extract, Transform and Load (ETL) applications. The serverless application will get some data, transform with a FaaS and store it in a database with BaaS. This pattern should be found at the edge in a distributed way. Small devices could collect and aggregate data acquired. Then data can be transformed and processed on the same device or an edge node and finally transferred to storage in the cloud. Data aggregation would reduce bandwidth costs to the cloud. Data are processed locally, and less data is transferred to the cloud to be stored or to be aggregate in a larger process involving many data from many places.

However, there are still challenges to be addressed. Security is essential but difficult to achieve. All interaction with BaaS increases the attack surface. It is also currently difficult to manage the hardware heterogeneity. Serverless at the edge should involve many different devices at every stage of fog computing, from the sensor to the cloud with cloudlet and edge nodes.

Sledge \cite{SledgeAServerlessFirstLightWeightWasmRuntimefortheEdge} is an open-source for ServerLess at the EDGE. It is designed to run on a single-host server; with Wasm support, it can run on most servers from the smallest like Raspberry Pi to the more powerful ones. To achieve ultra-low latency, Sledge bypasses the kernel scheduler and takes advantage of Wasm. This helps to have a rapid startup time. Sledge uses a custom aWams compiler and runtimes. Wasm also provides benefits in terms of security with process isolation and sandboxing. With Wasm, it is possible to execute the same binary on all platforms running Sledge.

Nuclio, OpenFaaS are other frameworks for serverless at the edge. Unlike Sledge, they are not based on Wasm but on container technology. They can run on multiple platforms and be scheduled by Kubernetes.

The Content Delivery Network (CDN) notion can be extended to Access Delivery Network (ADN). The idea is not only to provide content but also computation capabilities in the vicinity of the users.

WebAssembly could be a key enabler for serverless at the edge. WASM permit developers to stay focus on their application without worrying about available architectures. Then, every application could be deployed at the edge.

Serverless is an important paradigm for edge computing. Many use cases can be solved with FaaS, e.g., aggregating and processing data from IoT devices.

\subsection{Automation}

In order to save time and reduce cost, automation has a role to play.

\subsubsection{Software development and deployment}

Practices from DevOps workflow should be adapted to edge computing.

The applications should be agnostic of where there will be executed, and this should not affect the development. In addition, developers should be able to access hardware accelerators easily. A simple interface should be accessible to defines the needs of the applications. Then, the scheduler will find the best nodes that achieve these needs.

Operating an edge node should be as simple as a traditional node in the cloud, but there are significant architectural differences. An application executed at the edge needs to be able to run on the infrastructure of most providers; in some cases, due to the node location, the user will have only one choice of provider. Application deployment needs to be multi-cloud friendly. A seamless and provider agnostic deployment at the edge should be possible.

Techniques of Continuous Integration and Continuous Delivery (CI/CD) have to be adapted. The advent of a highly heterogeneous environment
raises the need for an execution standard. It will not be possible for a team to test every kind of architecture available on edge nodes to get continuous deployment.

GitOps, a set of practices based on version control and declarative specification (e.g., yaml files), permits the quick deployment of infrastructures. Edge infrastructures could take advantage of GitOps to facilitate their management.

\subsubsection{Infrastructure automation}

Edge computing needs infrastructure automation to be resilient. With GKE Autopilot~\cite{gke_autopilot}, Google proposes a "Kubernetes serverless". The idea is to automate the management nodes and pool of nodes, not only the pods. This improvement will bring flexibility and scalability for clusters.
Virtual clusters and virtual networking can make infrastructure deployment easier.
Automated infrastructure management and multi-cloud manager will help deal with structural heterogeneity in the edge cloud.

\section{Conclusion}

Most of the data will be processed at the edge in the following years. Proposing a seamless cloud experience is becoming more and more important and raises new challenges. 

Architectures have to be adapted to become more flexible. In response to the mobility of devices, topology needs to be dynamic and resilient. The topology should also reflect the heterogeneity of the devices at the edge. In addition, these devices have special needs in term of energy consumption. Networking is a keystone for a seamless cloud experience, this is essential for the proper functioning of the applications. Virtualisation and cloud-native technologies help improve networking, and networking technologies help edge computing deployment.

The major challenge in execution at the edge is being able to run applications on every node seamlessly without worrying about its platform or architecture. WebAssembly is one of the best candidates to manage this heterogeneity. With a strong community behind its development, current limitations should be solved soon. As long as CPUs remain limited in computation power and energy efficiency, hardware accelerator support is a priority. We believe that higher abstraction interfaces are a way to make accelerators available without the need to make them mandatory. It is also essential to have an overview of the needs of accelerators to permit efficient scheduling.

Orchestration of computation from device to data centre needs to rely on technologies similar to Kubernetes. Some improvements are needed to make Kubernetes truly ready for the edge: managing heterogeneous devices, working with non-reliable connection and handling node churn and resources volatility. Access to accelerators is also important to consider.

MEC allows joining telco networks and edge computing which makes it an important paradigm to consider, 5G is one of the significant edge computing use cases. With a seamless cloud experience, deploying application at the edge should be as easy as using serverless functions.

\bibliographystyle{IEEEtranS}
\bibliography{references}

%\listoftodos
\end{document}